\documentclass[twocolumn,showpacs,amssymb]{revtex4}
\usepackage{graphicx}
\usepackage{dcolumn}
\usepackage{bm}

\def\beq{\begin{equation}}
\def\enq{\end{equation}}
\def\ba{\begin{eqnarray}}
\def\ea{\end{eqnarray}}

\def\<{<\!\!}
\def\>{\!\!>}

\begin{document}
\input{epsf}

\title{Tidal disruption jets of supermassive black holes as hidden sources of cosmic rays: explaining the IceCube TeV-PeV neutrinos }

\author{Xiang-Yu Wang$^{1,2}$, Ruo-Yu Liu$^{1,2,3}$, }

\affiliation{School of Astronomy and Space Science, Nanjing University, Nanjing, 210093, China;\\
$^2$Key laboratory of Modern Astronomy and Astrophysics (Nanjing
University), Ministry of Education, Nanjing 210093, China\\
Max-Planck-Institut f\"ur Kernphysik, 69117 Heidelberg, Germany}

\begin{abstract}
Cosmic ray interactions that produce high-energy neutrinos also
inevitably generate high-energy gamma rays, which finally
contribute  to  the diffuse high-energy gamma-ray background after
they escape the sources. It was recently found that, the high flux
of neutrinos at $\sim30$ TeV detected by IceCube lead to a
cumulative gamma-ray flux  exceeding the Fermi isotropic gamma-ray
background at 10-100 GeV, implying that the neutrinos are produced
by hidden sources of cosmic rays, where GeV-TeV gamma-rays are not
transparent. Here we suggest that relativistic jets in tidal
disruption events (TDEs) of supermassive black holes are such
hidden sources. We consider the jet propagation in an extended,
optically thick envelope around the black hole, which is resulted
from the ejected material during the disruption. While powerful
jets can break free from the envelope, less powerful jets would be
choked inside the envelope. The jets accelerate cosmic rays
through internal shocks or reverse shocks and further produce
neutrinos via interaction with the surrounding dense medium or
photons. All three TDE jets discovered so far are not detected by
Fermi/LAT, suggesting that GeV-TeV gamma-rays are absorbed in
these jets. The cumulative neutrino flux from TDE jets can account
for the neutrino flux observed by IceCube at PeV energies and may
also account for the higher flux at $\sim30$ TeV if less powerful,
choked jets are present in the majority of  TDEs.

\end{abstract}

\pacs{95.85Ry, 98.70Qy, 98.70Sa}
\maketitle

{\em Introduction---} Extraterrestrial neutrinos has been detected
in various analyses and  found to be consistent with an isotropic
flux of neutrinos that is expected from extragalactic
astrophysical source populations \cite{IceCube}. The source of the
IceCube neutrinos is still controversial. The proposed
astrophysical sources include galaxies with intense star-formation
\cite{SF-model},  jets and/or cores of active galactic nuclei
(AGNs)\cite{AGN-model}, gamma-ray bursts
\cite{GRB-model,Murase2013b} and etc. A recent combined likelihood
analysis gives a best-fit power-law flux of
$\varepsilon_\nu\Phi_\nu=6.7\times10^{-8}(E/100{\rm
TeV})^{-0.5\pm0.09}{\rm GeV cm^{-2} s^{-1} sr^{-1}}$ (all-flavor)
in the energy range from  25 TeV to 2.8 PeV \cite{IceCube}. The
astrophysical high-energy neutrinos are generated in the decay of
charged pions produced in inelastic hadronuclear ($pp$) and/or
photohadronic ($p\gamma$) processes of cosmic rays (CRs), both of
which generate high-energy gamma-rays from the decay neutral
pions. As the combined analysis gives an all-flavor neutrino flux
of $10^{-7}{\rm GeV cm^{-2} s^{-1} sr^{-1}}$ at about 30 TeV, it
is argued that the cumulative gamma-ray flux associated with the
neutrino emission is in tension with the Fermi diffuse
extragalactic gamma-ray background (EGB) at 10-100 GeV
\cite{Murase2013,Murase2015}. The case gets stronger as new
studies of the EGB composition at energies above 50 GeV find a
dominant contribution from blazars,  leaving only a $\sim14\%$
residual component for all other sources classes \cite{Fermi2015}.
Motivated by this, it is argued that IceCube neutrinos may come
from CR accelerators that are hidden in GeV-TeV gamma-rays, so
they would not overshoot the diffuse EGB
\cite{Murase2015,Bechtol2015}. Choked jets in collapsing massive
stars \cite{choked-GRB} and  cores of active galactic nuclei have
been suggested to be such hidden sources
\cite{Murase2015,Murase2013b}. Here we propose a new hidden source
model, i.e relativistic jets in tidal disruption events (TDEs) of
supermassive black holes (SMBHs).

In TDEs, a star is torn apart by gravitational tidal forces of a
SMBH, leading to a transient accretion disk which produces  a
bright panchromatic flare\cite{TDF}. There are growing number of
candidate TDEs have been discovered in X-ray, ultraviolet  and
optical surveys (see \cite{Komossa2015} for a review). Three TDE
candidate have been also detected in non-thermal X-ray and radio
emission, i.e. Swift J1644+57, J2058+05 and J1112-8238
\cite{Burrows2011,Bloom2011,Zauderer2011,Cenko2012,Brown2015}. The
non-thermal X-ray and radio emissions are thought to be produced
by relativistic jets, in which shocks occur and accelerate
non-thermal electrons. These shocks  may also accelerate CR
protons \cite{Farrar2009, Wang2011,Farrar2014}, which can produce
neutrinos via interaction with X-ray photons. Recent studies
suggest the presence of an extended, quasi-spherical, optically
thick envelope around the SMBH in TDEs\cite{Loeb1997,
Guillochon2014,Coughlin2014,Roth2015}. Here we suggest that CRs
accelerated by jets as they are propagating in the envelope can
also produce neutrinos via interaction with surrounding dense
medium  and/or photons. Fermi and VERITAS observations of Sw
J1644+57 have failed to detect high-energy emission above 100 MeV
during the X-ray flare \cite{Burrows2011}. Analysis of the
Fermi/LAT data of the other two TDE jet flares also find no
high-energy emission \cite{Peng2015}, suggesting that TDE jet
flares are hidden sources in GeV-TeV gamma-rays.

{\em Jet propagation and dissipation---} The environment
surrounding the TDE jet, after its launch, may be complex and
needs detailed numerical works. It is thought that, an extended,
quasi-spherical, optically thick gas is present around the SMBH
after disruption\cite{Loeb1997,
Guillochon2014,Coughlin2014,wind,Roth2015}. The presence of the
gas can solve the puzzle that the temperatures (few $10^4$ K)
found in optically discovered TDEs are significantly lower than
the predicted thermal temperature ($>10^5{\rm K}$) of the
accretion disk\cite{TDE-reprocess}. The gas at large radii can
absorb UV photons produced by the inner accretion disk and
re-emits it at lower temperatures of a few times $10^4$ K. This
reprocessing region may be due to the formation of a
radiation-dominated envelope around the SMBH \cite{Loeb1997,
Guillochon2014,Coughlin2014}, or a super-Eddington wind outflow
\cite{wind}. We invoke the envelope scenario to describe the
density profile of the gas environment for the purpose of
calculating the dynamics of the jet propagating through it.  The
wind outflow scenario may have a different density profile, but
the essence of jet dynamics and the nature of an optically thick
gas remain unchanged\cite{Roth2015}. Since at distance much larger
than tidal disruption radius the rotation is dynamically
unimportant, the strong radiation pressure disperses the
marginally bound gas into a quasi-spherical configuration
\cite{Loeb1997, Guillochon2014,Coughlin2014}. The density profile
in the optically thick envelope can be described by
\cite{Loeb1997,De Colle}
\begin{equation}
\rho_e(r)=\frac{f M_*}{4\pi {\rm ln}(R_{out}/R_{in})r^3}
\end{equation}
where $M_*$ is the mass of the disrupted star and $f\simeq0.5$ is
the fraction of the mass in the envelope (which is the mass of the
stellar debris that is bound to the massive black hole), and
$R_{out}$ and $R_{in}$ are the outer and inner radii of the
envelope respectively. The inner radius is the tidal disruption
radius $R_{in}=R_t=R_*(M_{\rm
BH}/M_*)^{1/3}\simeq4\times10^{13}{\rm cm}(M_{\rm
BH}/10^7M_\odot)^{1/3}$, where $M_{\rm BH}$ is the mass of the
SMBH and $R_*\simeq R_\odot$ is the initial radius of the
disrupted star. $R_{out}$ is the radius where the envelope becomes
optically thin, which for electron scattering opacity is given by
\begin{equation}
 R_{out}=1.7\times10^{15}{\rm cm}(\frac{f
M_*}{0.5M_\odot})^{1/2}
\end{equation}
We use c.g.s. units and the denotation $Q=10^xQ_x$ throughout the
paper.

The propagation of a TDE jet in the extended envelope has been
studied with numerical simulations\cite{De Colle}. As the jet
advances in the surrounding envelope, the jet drives a bow shock
ahead of it. The jet is capped by a termination shock, and a
reverse shock propagates back into the jet, where the jet is
decelerated and heated. The jet head velocity is obtained by the
longitudinal balance between the momentum flux in the shocked jet
material and that of the shocked surrounding medium, measured in
the frame comoving with the advancing head \cite{jet-head}, which
gives
\begin{equation}
v_h=(\frac{L}{4\pi r^2 c \rho_e})^{1/2} =10^{10}{\rm cm s^{-1}}
L_{48}^{1/2}r_{15}^{1/2}.
\end{equation}
where $L$ is the isotropic luminosity of the jet, $\rho_e\simeq
2\times10^{-14} {\rm g \, cm^{-3}}r_{15}^{-3}$. While the jet is
propagating inside the envelope with a sub-relativistic velocity,
a significant fraction of jet "waste" energy is pumped into the
cocoon surrounding the advancing jet\cite{De Colle}. The jet can
break out of the envelope if the jet energy supply lasts longer
than the break-out time, which is
\begin{equation}
t_{br}=\int \frac{dr}{v_h}=2\times10^5{\rm s} \,
r_{15}^{1/2}L_{48}^{-1/2}.
\end{equation}

Assuming that the mass accretion rate follows the fallback time of
stellar material onto the central black hole, matter returns to
the region near the pericenter radius at a rate  $\dot{M}\propto
(t/t_p)^{-5/3}$. The characteristic timescale $t_p$ for initiation
of this power-law accretion rate is the orbital period for the
most bound matter, which is \cite{Lodato2009,Jet-power}
\begin{equation}
t_p=1.5\times10^6 {\rm s}\,
(\frac{M_*}{M_\odot})^{(1-3\xi)/2}(\frac{M_{\rm
BH}}{10^7{M_\odot}})^{1/2}
\end{equation}
for a radiative, main-sequence star being disrupted, where
$\xi\simeq0.2-0.4$ is a parameter characterizing the the
mass-radius relation (i.e. $R=R_\odot(M_*/M_\odot)^{1-\xi}$)
\cite{Kippenhahn1994}. As the jet power may scale with the
accretion rate as $L\propto\dot{M}$ in the super-Eddington
accretion phase\cite{Jet-power}, the  duration of the jet peak
luminosity is $t_p$, and  $L\propto (t/t_p)^{-5/3}$ after that.
Comparing this time with the jet break-out time in Eq. (4), we
find that powerful jets with $L\ge10^{46.5}M_{\rm BH,7}{\rm erg
s^{-1}}$ can break out the envelope successfully, while jets with
luminosity $L\le 10^{46.5}M_{\rm BH,7}{\rm erg s^{-1}}$  would be
choked in the envelope. In the latter case, all the jet energy is
transferred to the cocoon. If the energy accumulated in the cocoon
is larger than the binding energy of the outer part of the
envelope, which is about $10^{51}{\rm erg}$ for a $10^7M_\odot$
SMBH, the cocoon may unbind part of the envelope.

At the jet head, reverse shocks heat the jet material and
accelerate protons and electrons. Internal shocks may also arise
from the internal collisions within the jets resulted from the
fluctuations in the jet bulk Lorentz factor $\Gamma$
\cite{internal-shock}. It is thought that, the  variable X-ray
emission of Sw J1644 is produced by internal shocks. As the
observed minimum X-ray variability time is $t_v\simeq 100$ s
\cite{Burrows2011}, internal shocks may occur at a radius of
\begin{equation} R\simeq 2\Gamma^2 c
t_v=6\times10^{14} \Gamma_1^2 t_{v,2} {\rm cm}.
\end{equation}

TDEs with relativistic jets of $L>10^{48}{\rm erg s^{-1}}$ imply
accretion rates $>10^3$ higher than the Eddington accretion rate for
$10^7 M_\odot$ black holes. These jets can break free from the
optically-thick envelopes, and become optically thin after
they cleared open channels. There may be less powerful
relativistic jets with $L<10^{48}{\rm erg s^{-1}}$, as long as the
accretion is super-Eddington. As the bulk Lorentz factor may also
be lower for a less powerful TDE jet, the dissipation of the jet
energy may occur at radius smaller than $10^{15}{\rm cm}$ for
internal shocks, well within the optically thick envelope. When
$L<10^{46}{\rm erg s^{-1}}$, the jet will be choked and all the
dissipation processes can only occur inside the optically thick
envelope.

{\em Neutrino production ---} We assume that the composition of
the jet is mainly protons and electrons. Although the composition
of the nascent jet produced from the central black hole is unknown
and could be magnetically dominated. However, as the jet burrows
through the surrounding gas, protons from the surroundings could
be entrained into the jet. According to numerical simulations of
jet propagation \cite{Zhang2013}, Kelvin-Helmholtz instabilities
and/or oblique shocks that develop lead to the mixing of
surrounding material into the jet  while the jet is advancing with
a sub-relativistic velocity.

Internal shocks and reverse shocks that propagate into the
low-density jets are collisionless, although they locate inside
the optically thick envelope. CR acceleration is expected as the
shocks are not affected by the radiation. This is because the
mean-free-path of thermal photons propagating into the upstream
flow $l_{\rm dec}=(n_j\sigma_{\rm T})^{-1}=10^{17}{\rm
cm}L_{48}^{-1}r_{15}^2\Gamma_{1}^2$ is much larger than the
comoving size of the upstream flow $l_u=r/\Gamma=10^{14}{\rm
cm}r_{15}\Gamma_{1}^{-1}$ \cite{Murase2013b}, where $n_j$ is the
upstream proton density. It has been shown that shocks in TDE jets
can accelerate cosmic rays to ultra high energies
\cite{Farrar2009, Wang2011, Farrar2014}.

We first consider the neutrinos produced by high-luminosity jets,
which can successfully break free from the envelope. The jet
clears the material in the channel during the breakout and the
radiation from the jet can escape after the jet breaks out.
Luminous non-thermal X-ray emission has been seen in three such
TDE jets. The X-ray emission should be produced by relativistic
electrons accelerated in the jets. The CR protons may interact
with these non-thermal X-ray photons and produce neutrinos. The
effective $p\gamma$ efficiency, defined as the ratio between the
dynamic time and the $p\gamma$ cooling time ($f_{p\gamma}\equiv
t_{\rm dyn}/t_{p\gamma}$), is
\begin{equation}
f_{p\gamma}(\varepsilon_p)=\sigma_{p\gamma}n'_{X}(r/\Gamma)K_{p\gamma}\simeq2
L_{\rm X,48} \Gamma_1^{-2} r_{15}^{-1}\epsilon_{\rm X, KeV}^{-1}
\end{equation}
where $\sigma_{p\gamma}\simeq5\times10^{-28}{\rm cm^2}$ is the
peak cross section at the $\Delta$ resonance, $n'_{X}$ is the
number density of X-ray photons in the comoving frame of the
shock, $\varepsilon_{p}=0.15 {\rm
GeV^2}\Gamma^2/\epsilon_X=1.5\times10^{16}\Gamma_1^2\epsilon_{\rm
X,KeV}^{-1} {\rm eV}$ is the proton  energy that interact with
X-ray photons and  $\varepsilon_{\nu }\simeq
7.5\times10^{14}\Gamma_1^2\epsilon_{\rm X,KeV}^{-1} {\rm eV}$ is
the corresponding neutrino energy. Such a high efficiency for
$p\gamma$ interactions can   be in fact inferred from the
gamma-ray opacity in  three detected TDE jets. Analyses of the
Fermi/LAT data of all three jetted TDEs  find that they are not
detected by Fermi/LAT, with a flux limit in LAT energy being less
than $1\%$ of the flux in X-rays (Peng et al. 2015). The
non-detection of high-energy gamma-rays is most likely due to that
the emitting region is not transparent to these gamma-rays, i.e.
$\tau_{\gamma\gamma}(\varepsilon_\gamma\ge100 {\rm MeV})>1$. It is
useful to express $f_{p\gamma}$ as a function of the pair
production optical depth $\tau_{\gamma\gamma}$. The optical depth
for pair production of a photon of energy $\varepsilon_h=100{\rm
MeV}$ is
$\tau_{\gamma\gamma}(\epsilon_h)=\sigma_{\gamma\gamma}n'({\epsilon_t})(R/\Gamma)$,
where $n'(\epsilon_t)$ is the number density of target photons,
which have an energy of $\epsilon_t=(\Gamma m_e
c^2)^2/\epsilon_h=250{\rm KeV}(\Gamma/10)^2$. For a power-law
spectrum $\beta=2$ for target photons, we have
$f_{p\gamma}(\varepsilon_\nu=750{\rm
TeV})\simeq0.5\tau_{\gamma\gamma}(100{\rm MeV})$. So we reach the
conclusion that the  pion production efficiency is high  for the
high-luminosity jets.

Now consider the neutrino emission produced by low-luminosity,
choked jets. CR protons can produce neutrinos by colliding with
dense gas in the optically thick envelope.   The effective pion
production efficiency for $pp$ collisions is
\begin{equation}
f_{pp}=\sigma_{pp}(\rho_e/m_p) r K_{pp}=20(r/10^{14}{\rm
cm})^{-2}.
\end{equation}
where $\sigma_{pp}\simeq 4\times10^{-26}{\rm cm^2}$ is the cross
section of $pp$ interaction and $K_{pp}\simeq 0.5$ is the
inelasticity. For low-luminosity jets, $\Gamma$ may be lower (e.g.
$\Gamma\simeq 3$), then internal shocks may occur at radius
$<10^{14.5}{\rm cm}$ so that $f_{pp}> 1$. Internal shocks may also
produce non-thermal X-rays as has been seen in three jetted TDEs.
Then, the effective $p\gamma$ efficiency is
\begin{equation}
f_{p\gamma}(\varepsilon_p)\simeq2 L_{\rm X,46} \Gamma_{0.5}^{-2}
r_{14}^{-1}\epsilon_{\rm X, KeV}^{-1}
\end{equation}
where $\varepsilon_{p}=1.5\times10^{15}{\rm eV}
\Gamma_{0.5}^2\epsilon_{\rm X,KeV}^{-1}$ and the neutrino energy
is $\varepsilon_{\nu }\simeq 75 {\rm TeV}
\Gamma_{0.5}^2\epsilon_{\rm X,KeV}^{-1}$.

The envelope may  contain dense thermal photons\citep{Loeb1997},
then $p\gamma$ interaction with these thermal photons would also
be important. According to \citep{Loeb1997}, the effective
temperature at the photosphere $R_{out}$ is $T_{p}=2\times10^4{\rm
K}(M_{BH}/10^7M_\odot)^{1/4}$ and the interior temperature scales
as $T\propto \rho^{1/3}\propto r^{-1}$, so $T=4\times10^4 {\rm
K}(r/10^{15}{\rm cm})^{-1}$. The corresponding photon number
density is $n_\gamma=2\times10^{14}{\rm cm^{-3}}(r/10^{15}{\rm
cm})^{-3}$. The CR protons interacting with these soft photons
have a typical energy of $\varepsilon_p=0.15{\rm
GeV^2}/(kT)=4\times10^{16}{\rm eV}$, where $kT$ is the typical
energy of thermal photons in the envelope. The corresponding
neutrino energy is $\varepsilon_\nu=0.05\varepsilon_p=2{\rm PeV}$.
The effective pion production efficiency for $p\gamma$ collisions
is
\begin{equation}
f_{p\gamma}=\sigma_{p\gamma}n_\gamma r
K_{p\gamma}=20(r/10^{15}{\rm cm})^{-2}.
\end{equation}

For these choked TDE jets, as the neutrino production site is
within the optically thick region, the associated high-energy
gamma-rays can not escape. Instead, high-energy gamma-rays are
absorbed by low-energy electrons and photons in the envelope,
depositing their energy finally into the envelope. Therefore,
these choked TDE jets are also hidden sources of CRs.

{\em CR and neutrino flux---}  We now estimate the CR and neutrino
flux produced by TDE jets. The three TDEs with relativistic jets
detected by Swift imply a local rate of $0.03 {\rm Gpc^{-3}
yr^{-1}}$ for  jet luminosity  $L\ge10^{48}{\rm erg s^{-1}}$
\cite{Farrar2014,Sun2015}. The isotropic radiation energies in
X-rays in all three jetted TDEs are about $3\times10^{53}{\rm
erg}$ \cite{Burrows2011,Cenko2012}. Assuming that the total
bolometric radiation energy is three times larger, the energy
injection rate in radiation is about $3\times 10^{43}{\rm erg
Mpc^{-3} yr^{-1}}$. If the electrons occupy a fraction of
$\epsilon_e=0.1$ of the proton energy, then the energy injection
rate in protons is about $\dot{W}_{p,z=0}=3\times 10^{44}{\rm erg
Mpc^{-3} yr^{-1}}$. From the kinetic energy of the jet obtained
with the radio modeling and assuming a beam factor of
$f_b=10^{-3}$ for relativistic jets, Farrar \& Piran
\cite{Farrar2014} obtained a similar energy injection rate,
$2\times 10^{44}(f_b/10^{-3})^{-1}{\rm erg Mpc^{-3} yr^{-1}}$.
This rate is roughly what is needed by the flux of ultra-high
energy cosmic rays (UHECRs), and Farrar \& Piran \cite{Farrar2014}
suggested that tidal disruption jets may be the source of UHECRs.

The above estimate does not include the contribution by  less
powerful ($L<10^{48}{\rm erg s^{-1}}$) or even choked
($L<10^{46.5}{\rm erg s^{-1}}$) jets that may be present in normal
TDEs.  If the jetted TDEs detected by Swift follow the
extrapolation of normal TDE luminosity function to
high-luminosities \cite{Sun2015},  we would expect that there are
more TDEs harbouring jets with luminosity of
$\sim10^{46}-10^{47}{\rm erg s^{-1}}$. Since the peak accretion
rate in TDEs is generally super-Eddington, jet formation  is
naturally expected in all TDEs. We assume that the majority of
normal TDEs have a low-luminosity jet with $L\sim10^{46}{\rm erg
s^{-1}}$, so that they are choked and do not have bright radio
afterglow emission, consistent with radio observations of normal
TDEs\cite{Radio}. Assuming a peak accretion time given by Eq. (5),
the total energy in the choked jet would be $\simeq 10^{51}
f_{b,-1}{\rm erg }$, where $f_b\simeq 0.1$ is the beam correction
fraction.  The event rate of normal TDEs is as high as $10^{3}{\rm
Gpc^{-3}yr^{-1}}$\cite{rate},  so the energy injection rate by
choked jet could be as large as $10^{45}{\rm erg Mpc^{-3}
yr^{-1}}$.

To account for the IceCube neutrino flux, the required local
energy injection rate  is
\begin{equation}
\begin{array}{ll}
{\dot W_{p,z=0}}=\alpha\left(\frac{c \xi_z}{4\pi H_0}\right)^{-1}
\varepsilon_p \Phi_p \\
=10^{45} {\rm erg Mpc^{-3}
yr^{-1}}\left(\frac{\alpha}{10}\right)f_{\pi}^{-1}\left(\frac{\xi_z}{3}\right)^{-1}\left(\frac{\varepsilon_\nu\Phi_\nu}{10^{-7}{\rm
GeV cm^{-2} s^{-1} sr^{-1}}}\right)
\end{array}
\end{equation}
where $\varepsilon_p\Phi_p$ is the  proton  flux, $\alpha$ is a
factor coming from normalization of the proton spectrum (e.g., for
a power-law index $s=2$ of the cosmic ray spectrum, $\alpha={\rm
ln}(\varepsilon_{p,max}/\varepsilon_{p,min})$), $\xi_z$ is a
factor accounting for the contribution from high-redshift sources
and $\varepsilon_\nu\Phi_\nu$ is the all-flavor neutrino flux.
Here $f_{\pi}\equiv1-{\rm exp}(-f_{p\gamma, pp})$ is the fraction
of proton energy lost to pions through $p\gamma$ or $pp$
collision. Since $f_\pi \simeq1$ for our case, the energy
injection rate by TDEs can account for the neutrino flux observed
by IceCube at PeV energies and may even account for the higher
flux at $\sim30$ TeV if less powerful, choked  jets from normal
TDEs are included.

{\em Discussions } Detection of neutrinos from one single TDE jet
with ${\rm KM^3}$-scale neutrino detector requires that the source
must be extremely bright with a total electromagnetic fluence
$\ge10^{-3}{\rm erg cm^{-2}}$. Among the three TDEs with
relativistic jets, only Sw J1644+57 has such a large fluence.
Thus, the stacking search of a dozen of jetted TDEs is needed to
fulfill a promising detection. For choked jets that have an
isotropic equivalent energy of $10^{52}{\rm erg}$,  detection of
one neutrino requires that the TDE should be at distance within
200 Mpc and that the jet  point to the observer. Considering the
beam fraction $f_b=0.1$ of the jet and a TDE rate of $10^3{\rm
Gpc^{-3} yr^{-1}}$, the number of such TDEs in the observable
volume  is only $N=3f_{b,0.1}$ per year. Thus, stacking search for
neutrinos is also needed for nearby normal TDEs with choked jets.

This work is supported by the 973 program under grant
2014CB845800, the NSFC under grant 11273016, and the  Excellent
Youth  Foundation  of  Jiangsu Province (BK2012011).

\end{document}